# Influence of surface termination on inverse Goos—Hänchen shift of negatively refractive photonic crystals

Jinbing Hu[1,2*], Binming Liang[1], Jiabi Chen[1], Xiaoshu Cai[2], QiangJiang[1], Songlin Zhuang[1]

[1]Shanghai Key Lab of Modern Optical System, University of Shanghai for Science and Technology, No. 516 Jun Gong Road, Shanghai 200093, China

[2]Institute of Particle and Two Phase Flow Measurement, School of Energy and Power Engineering, University of Shanghai for Science and Technology, No.516 Jun Gong Road, Shanghai 200093, China.

*Corresponding authors: hujinbing@usst.edu.cn

The effect of surface termination on the inverse Goos—Hänchen (GH) shift of two-dimensional (2D) negatively refractive photonic crystal (NRPhC) containing air holes arranged in hexagonal lattice in a dielectric background is investigated for TM polarization. Results show that the magnitude of the inverse GH shift of 2D-NRPhC strongly depends on surface termination even for an incident beam with a fixed frequency and incidence angle. Further study by calculating the dispersion of surface mode of 2D-NRPhC as a function of surface termination reveals that 2D-NRPhC presents large inverse Goos-Hänchen shift at those terminations where surface mode is excited; thus, large inverse Goos-Hänchen shift originates from backward surface mode of 2D-NRPhC. In addition, the coupling coefficient of incident field into the field of surface mode as a function of surface termination is studied and  demonstrates above results. This paper provides technical information regarding the combination of various functional photonic elements in the design of integrated optical circuits.

## 1. Introduction

As early as 1968, Veselago[1] predicted that electromagnetic waves that travel through materials with negative index of refraction would result in opposite effects, such as reversed Doppler Effect, and reversed Vavilov—Cerenkov effect, et al. And it was reported[2, 3] that if the so-called equifrequency surface (EFS) of photonic crystal (PhC) satisfies certain condition PhC mimics the behavior of dielectric materials with a negative index of refraction in certain spectral regions. The subject of negative refraction in NRTPhC has inspired a lot of interest in physics for its important potential applications. Most recently, our group[4] experimentally observed the inverse Doppler Effect in negatively refractive PhC (NRPhC) at optical frequency; this study was an important breakthrough in the study of PhC and provided a basis for further exploration of exceptional phenomena in 2D-NRPhC[5]. Another exceptional phenomenon regarding NRPhC is the inverse Goos-Hänchen effect [6,7] occurring at the interface between normal material (e.g., air) and NRPhC. As is well known, when a bounded wave beam is incident from an optically denser medium to a medium less optical dense at critical angle, the centre of reflected beam undergoes a sizable lateral shift from the position predicted by geometrical optics: This is the Goos-Hänchen (GH) shift [8]. The inverse GH shift refers to a shift on the same side of the normal as the incident beam [6]. The GH shift occurs as a result of the different phase acquisition of the plane wave components of the finite incident beam and is given as $S = -d\phi / kd\theta$ [9]. $\phi$ is the phase of reflected beam. $k$ and $\theta$ are the wave vector number and the incidence angle, respectively, in the incident medium.

It has been reported in papers [2, 10] that PhC under certain conditions can behave refractive as obeying Snell's law with a negative effective phase index $|n_p| < 1$. Thus wave, propagating from air to the PhC, is essentially wave coming from an optically denser medium to a less dense optical medium; hence inverse GH shift will take place when the incidence angle equals to or even above the critical angle. However, the physics of GH shift in PhC is a bit intricate due to the remarkable surface effect, because PhC is a periodic medium and the modes repeat periodically in the reciprocal space [11]. In addition, it was found that the properties of PhC are strongly influenced by the way the periodic PhC is terminated. The inverse GH effect, therefore, may be related to surface termination of NRPhC.

This paper aims to investigate the influence of surface termination on the inverse GH shift of NRPhC. To the best of our knowledge, no study has focused on this problem but few works have investigated the effect of surface termination on the optical properties of PhC [12-16]. Previous studies have investigated the role of surface termination on the reflection and transmission spectra of 2D-PhC; results have demonstrated that surface termination generates dips within the photonic stop bands of the reflection spectra in TM and TE polarizations[12, 13]. Lu et al.,[14] reported that electromagnetic waves can be guided at the edge of a three-dimensional PhC in air by using a specific design of the intersection of two surface planes. Another study has demonstrated that the beaming direction from a PhC waveguide can be controlled by changing its surface termination[15].

In the present paper we choose the structure with hexagonal air holes in dielectric slab as the 2D-NRPhC because such structure can be easily fabricated with negative effective index $|n_p| < 1$ in the second band for TM polarization [10]. By changing surface termination (position of surface plane) we calculate the inverse GH shift of 2D-NRPhC as a function of surface termination. Then, the dispersion of surface mode as a function of surface termination is given to understand the origin of the effect of surface termination on inverse GH shift. Last, by introducing the coupling coefficient of the incident field into the field of surface mode the characteristic of inverse GH shift is well explained.

In our calculation the plane wave expansion method (PWEM) [17] is used to obtain the frequency band structure and the equifrequency surface (EFS) of photonic crystal. And the supercell method [11] based on PWEM is used to get the dispersion of surface mode (as a function of surface termination) by applying the Bloch theorem. To get the curve of inverse GH shift of 2D-NRPhC as a function of surface termination we use commercial software *FDTD Solution* [18] with the boundary condition of PML [19] to simulate the reflection of bounded wave beam at the interface of air and NRPhC, and monitored the profile of reflected beam, then, the magnitude of inverse GH shift is extracted by obtaining the horizontal coordinate of the maximum value of reflected beam [20].

## 2. Structure

We consider a two-dimensional photonic crystal (2D-PhC), which consists of hexagonal lattice of air holes in a dielectric background with dielectric constant of 12.96 (e.g., GaAs at near infrared region). The radius of air holes $r = 0.4a$, where $a$ is the lattice constant. Here, we only take into account TM polarization. We set the $xz$ plane as the plane of periodicity, which is parallel to the propagation vector. The interface between air and 2D-PhC is parallel to the $\Gamma - K$ direction. Before we precede in present the sign of effective phase index and EFS of 2D-PhC it is helpful to discuss definition of the phase ($n_p$) and group ($n_g$) refractive indices and their associated phase ($v_p$) and group ($v_g$) velocities for infinite PhC system. Note that all the following discussion is restricted to zeroth-order Bragg beams, i.e., specular reflected beam and transmitted beam. For any general case, $v_p = \dfrac{c}{|n_p|}\hat{k}$, with $\hat{k} = k/k$. Also $|v_g| = |\nabla_k \omega| = \dfrac{c}{|n_g|}$. It has been proven analytically [21] that for the infinite PhC system the group velocity has the same direction as the energy velocity, i.e., Poynting vector $p$. Therefore, the sign of $v_g \cdot k$ is equivalent to the sign of $p \cdot k$, and the sign of $n_p$ is same as the sign of $v_g \cdot k$. Meanwhile, the sign of $n_p$ in PhC can be determined from the behavior of EFS [22]. If the equifrequency contours move outwards with increasing frequency then $v_g \cdot k > 0$; if the equifrequency contours move inwards $v_g \cdot k < 0$. The value of $n_p$ for a certain incidence angle $\theta$ will be $c|k_f(\theta)|/\omega$. For the special case of normal dispersion, the value of $n_p$ is independent of the incidence angle.

There are two prerequisites that should be satisfied for designing PhC with negative effective phase index [23]. First, the slope of band curve should be negative, meaning that the angle between the direction of group velocity and that of phase velocity is obtuse angle, i.e., $v_g \cdot k < 0$. Second, the band curve should be as symmetrical about $\Gamma$ as possible, implying that the EFS of this band are almost circular. Then, the PhC in the corresponding frequency region can be taken as isotropic material. With this in mind, we figure out the first four band of TM polarization (in Fig. 1(a)) and find that the second band (denoted as bole blue curve) satisfies that two prerequisites. Then, we plot the EFS of the second band curve in Fig. 1(b). The hexagon composed by dotted lines is the 1st Brillouin Zone (BZ). It is apparent that the EFS in the normalized frequency region from 0.3~0.34 $\omega a/2\pi c$ are circular, implying that the PhC in this frequency region is characterized by isotropy. In addition, the EFS shrink with increase of frequency, i.e., $v_g \cdot k < 0$. In other words, the direction of group velocity is opposite to that of phase velocity. To demonstrate this relation, we figure the directions of group and phase velocity in PhC when a beam encounters the surface of PhC from air according to Huygens principle, shown in Fig. 2. From these EFS [23] we can obtain the effective phase index of PhC, $n_p = -0.485$, corresponding to critical angle 29 deg. In Fig. 2(a), the beam with normalized frequency 0.33 is incident from air (left region) onto the surface (denoted by the black vertical dotted line) of PhC (right region) at critical angle, as shown by blue arrow. The brown horizontal dotted line is the conserved k line as well as the normal of surface. It is apparent that the group velocity (represented by green arrow) in PhC is opposite to the phase velocity (shown by red arrow). What's more, the group velocity is along the surface of PhC, i.e., total reflection. For incidence angle above critical angle, total reflection takes place, and inverse GH shift is possible. Fig. 2(b) shows the situation of incidence angle equaling to 30 deg. Note that to obtain one reflected and/or refracted beam, the neighboring equifrequency circles should not intersect with the conserved k line, as the situation shown in Fig. 2(b).

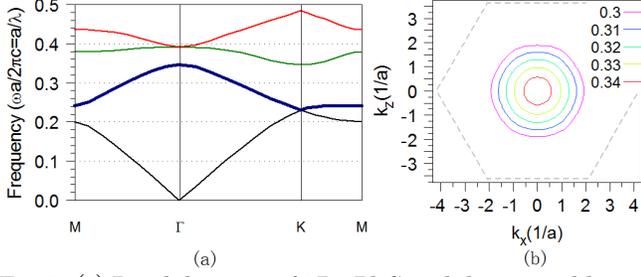

Fig. 1. (a) Band diagram of 2D –PhC with hexagonal lattice of air holes in dielectric slab, whose permittivity is 12.96. The radius of air holes $r = 0.4\ a$, where $a$ is the lattice constant. (b) The EFS of the second band (bold blue curve in Fig. 1(a)) in the 1st BZ (the dotted hexagon), the frequencies are uniform between 0.3~0.34 $\omega a / 2\pi c$ from the outermost curve to the innermost one.

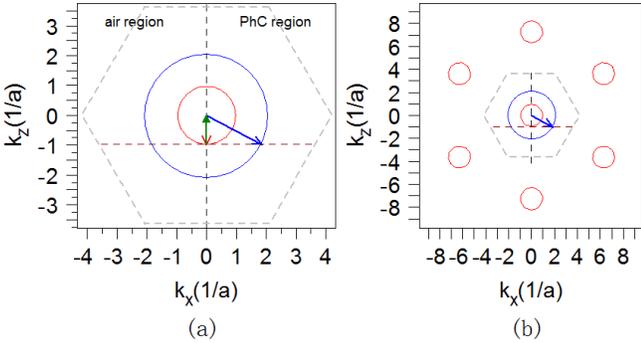

Fig. 2. The directions of group (denoted by green arrow) and phase (denoted by red arrow) velocity in (a) 1st BZ and (b) repeated zone for normalized frequency 0.33 red circle. The blue circle is the equifrequency contour of air. The blue arrow represents the incident beam, which is incident from air (left region) onto the surface (the black vertical dotted line) of PhC (right region) at (a) 29 deg and (b) 30 deg. The brown horizontal dotted line is the conserved k line as well as the normal of surface.

## 3. Results and Discussion

As been mentioned above that GH shift may occur when the bounded wave beam radiates (from air) onto the surface of NRPhC with effective index of refraction $|n_p| < 1$. However, the effect of surface termination on the GH shift of NRPhC has not been studied. Here, we explore the influence of surface termination on the inverse GH shift of NRPhC by means of commercial software- *FDTD Solution* [18]- when TM-polarized beam incidents the PhC slab (thick enough to mimic semi-infinite PhC) at critical angle. First of all, we introduce the termination parameter $\tau$ ($0 \le \tau < 1$), which corresponds to the length of the unit cell in $z$ direction and describes the surface termination. When $\tau = 0$, the outermost row of air holes exhibit a complete profile (i.e., perfect termination, as shown in the inset of Fig. 3) and linearly increases with increasing the height of the surface. The incident beam used to illuminate the interface of air and 2D-NRPhC has Gaussian profile in its cross-section with width of 16 $a$.

The results are presented in Fig. 3, in which the horizontal axis is the termination parameter $\tau$ and the vertical axis is the inverse GH shift. The result shows that, as expected, there is inverse and large GH shift at the surface of NRPhC when total reflection occurs. Unlike the normal GH shift of totally internal reflection of isotropic homogeneous materials, however, the inverse GH shift remains not constant on the condition that the beam is always critically incident, but is dramatically affected by surface termination. For surface termination $\tau = 0.25$ and 0.85 or so, the inverse GH shift gets its maximum values, $-28.27\ a$ and $-9.215\ a$, respectively. For other surface terminations the inverse GH shift of NRPhC is very small, even zero.

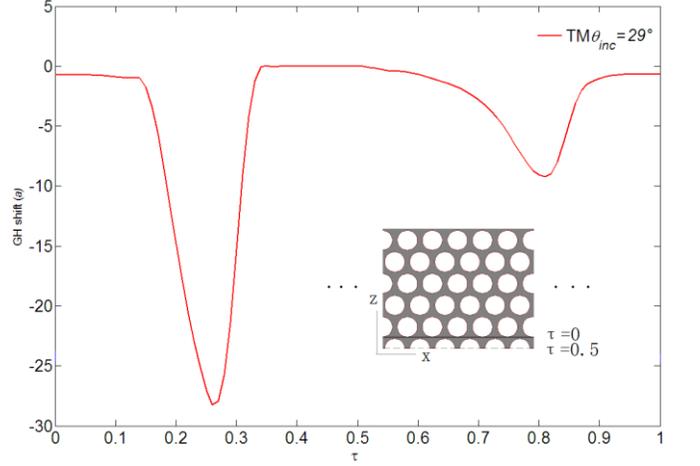

Fig. 3. The curve of inverse GH shift of 2D-NRPhC as a function of surface termination $\tau$ ($0 \le \tau < 1$, the inset shows the situation of $\tau = 0$ and 0.5), corresponding to 0.866 $a$, the period in $z$ direction. The beam incident at critical angle 29 deg has the characteristic of TM polarization.

The physical reason beneath this phenomenon may be as follow: For that frequency of incident beam (i.e., TM polarization with normalized frequency of 0.33) the PhC can be taken as isotropic material with negative index $|n_p| < 1$, thus wave, incident from air onto the surface of NRPhC, will be totally reflected when incidence angle equals to critical angle, and GH shift may take place. According to the electromagnetic variational principle [24], the low-frequency modes concentrate their energy in the high-$\varepsilon$ regions, and the high-frequency modes have a larger fraction of their energy in the low-$\varepsilon$ regions. With this in mind, we can know that with the increase of surface termination (i.e., the augment of high-$\varepsilon$ region) the frequency of surface modes is pulled down from the second band (also defined as air band) [25]. In other words, the slope of surface mode is negative, i.e., backward surface wave. Meanwhile, reports for PhC of air holes in dielectric slab have shown that the surface mode can exist on the surface of PhC and that the surface mode is very sensitive to the surface termination [26]. The combination of total reflection and surface effect are that those, transformed into surface wave of NRPhC when total reflection takes place, can radiate back into vacuum (just as in the case of surface waves in corrugated metal surfaces [27]) when propagates along the surface of PhC, hence inverse GH shift is observed for certain terminations (e.g., 0.25 or 0.85).

To confirm above mechanism we investigate the physical relation between the dispersion of surface mode and the surface termination τ with incidence angle fixed at critical angle. Due to the breaking of the translational symmetry in the normal direction we must consider a supercell that spans the entire PhC slab in the normal direction [11]. This supercell consists of only one unit cell along the lateral direction, where translational symmetry still applies. Nevertheless, several PhC sites (including the air region) should be taken along the normal direction. The number of sites, as well as the length of the air space on the top and bottom of the terminated PhC sites, must be thick enough to disallow any mode coupling between the top and bottom PhC slab. In our calculation the PhC slab has the thickness of 51 periods and the thickness of air regions on the top and bottom of the PhC are 7 period.

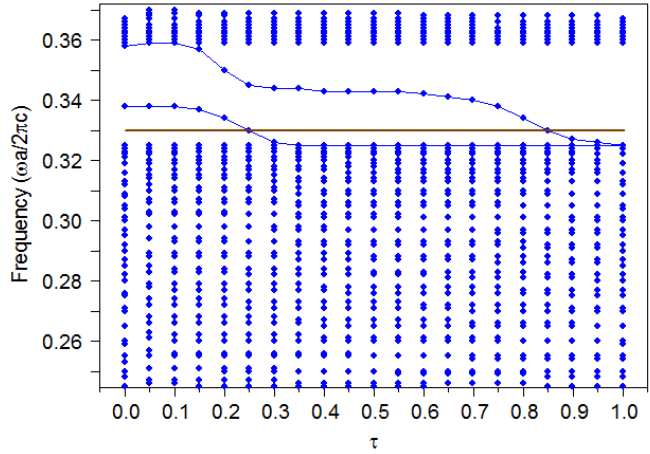

Fig. 4 The curve of the dispersion of surface mode with change of surface termination, where the parallel component of wave vector $k_x$ is 0.16 $2\pi / a$ .

The dispersion of surface mode as a function of surface termination is shown in Fig. 4, which shows that the dispersion of surface mode is very sensitive to the surface termination. From this figure we can observe that the line of normalized frequency 0.33 (denoted by the horizontal brown line) intersects with the dispersion of surface mode at terminations τ =0.25 and 0.85, which means that the incidence of 0.33 TM beam onto PhC with surface termination of 0.25 or 0.85 will excite surface wave at the surface of PhC. The results agree very well with that of Fig. 3, in which the inverse GH shift of normalized frequency of 0.33 TM polarization gets maxima value at terminations τ =0.25 and 0.85. Note that the frequency of surface mode indeed descends with increase of surface termination. In other words, the slope is negative and the surface wave is backward type.

Someone may find that why the inverse GH shifts of NRPhC with terminations τ = 0.25 and 0.85 are not same now that the surface modes are excited at those two terminations. To answer this question, we have to quantitatively study the relation between inverse GH shift and surface termination; we cannot obtain the reason from above researches (e.g., Figs 3 and 4) because they are qualitative. Here, we determine the relation between the magnitude of inverse GH shift and surface termination by introducing the wave coupling coefficient of incident field into the field of surface mode of NRPhC, because the more energy incident field coupled into surface wave the larger the magnitude of inverse GH shift. The definition of coupling coefficient is defined as the following formula [28]:

$$c = \frac{\langle S_{NRPhC} \rangle}{\langle S_{in} \rangle} \qquad (1)$$

where $\langle S_{NRPhC} \rangle$ and $\langle S_{in} \rangle$ represent the time-averaged power fluxes of the backward surface field in 2D-NRPhC and the incident field, respectively, within unit cell along the surface. The readers can find a complete description about the definition of coupling coefficient in Refs. 29 and 30.

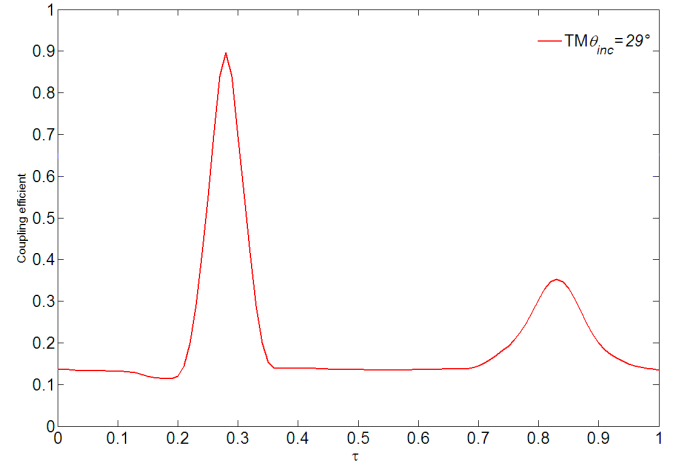

Fig.5. Curves of coupling efficient between incident field and the field of backward surface wave in 2D-NRPhC versus surface termination for TM polarizations. The beam is incident at the critical angle.

The curve of coupling efficient versus surface termination is presented in Fig. 5, which clearly show that the coupling coefficient varies with the change of surface termination. The coupling coefficient exhibits maximum values at τ = 0.25 and 0.85 or so due to the excitation of surface wave in this two terminations, and the value somewhere else is almost zero. Meanwhile, the coupling coefficient of

$\tau = 0.85$ is smaller than that of $\tau = 0.25$. Combined with Figs. 3 and 4, we can conclude that the inverse GH shift is resulted from backward surface wave, and the larger coupling coefficient (i.e., the more energy the incident field transformed into surface wave of NRPhC), the larger the inverse GH shift.

## 4. Conclusion

In conclusion, we have investigated the effect of surface termination on the inverse GH shift of 2D-NRPhC. Results show that surface termination plays a crucial role in the magnitude of inverse GH shift as well as the excitation of surface wave of 2D-NRPhC even for a beam with a fixed frequency and incidence angle. The study about the dispersion of surface mode of NRPhC as a function of surface termination reveals that the origin of large inverse GH shift is the excitation of backward surface mode at the surface of NRPhC, which produces inverse GH shift by radiating energy back into vacuum as they propagates along the interface of air and NRPhC. And this mechanism is proven by the coupling coefficient of incident field into backward surface field of 2D-NRPhC. The result of this paper will provide technically significant information regarding the combination of various functional photonic elements in the design of integrated optical circuits.

This work is partially supported by National Basic Research Program of China (2011CB707504), National Natural Science Foundation of China (61177043), The Innovation Fund Project for Graduate Student of Shanghai (JWCXSL1401).

Reference
[1] V. G. Veselago, "THE ELECTRODYNAMICS OF SUBSTANCES WITH SIMULTANEOUSLY NEGATIVE VALUES OF IMG align= ABSMIDDLE alt= ε eps/IMG AND μ," Physics-Uspekhi, vol. 10, pp. 509-514, 1968.
[2] S. Foteinopoulou and C. M. Soukoulis, "Negative refraction and left-handed behavior in two-dimensional photonic crystals," Physical Review B, vol. 67, p. 235107, 06/19/ 2003.
[3] Z.-Y. Li and K.-M. Ho, "Light propagation in semi-infinite photonic crystals and related waveguide structures," Physical Review B, vol. 68, p. 155101, 2003.
[4] J. Chen, Y. Wang, B. Jia, T. Geng, X. Li, L. Feng, W. Qian, B. Liang, X. Zhang, M. Gu, "Observation of the inverse Doppler effect in negative-index materials at optical frequencies," Nature Photonics, vol. 5, pp. 239-245, 2011.
[5] E. J. Reed, "Physical optics: Backwards Doppler shifts," Nature Photonics, vol. 5, pp. 199-200, 2011.
[6] L.-G. Wang and S.-Y. Zhu, "Large negative lateral shifts from the Kretschmann–Raether configuration with left-handed materials," Applied Physics Letters, vol. 87, pp. -, 2005.

[7]X. Wang, An. Jiang and F. Zheng, "large and bistable Goos-Hänchen shifts from the Kretschmann configureation with a nonlinear negaive -zero-positive index metammaterial," Journal of Optics, vol. 16, no. 4 pp.045101,2014.
[8] Goos. F and Hänchen. H, "Ein neuer und fundamentaler
versuch zur totalreflexion," Annalen der Physik, vol.436, no.7 pp.333-346, 1947.
[9] Artmann. K, "Berechnung der sei tenversetzung des totalreflektierten strahles," Annalen der Physik, vol.437, no.1 pp.87-102, 1948.
[10] S. Foteinopoulou, "Photonic crystals as metamaterials,"
Physica B: Condensed Matter, vol.407, no.20 pp.4056-4061, 2012.
[11] F. Ramos-Mendieta and P. Halevi, "Surface electromagnetic waves in two-dimensional photonic crystals: Effect of the position of the surface plane," Physical Review B,
vol.59, no.23 pp.15112–15120, 1999
[12] J. L. Garcia-Pomar, J. Gollub, J. Mock, D. Smith, and M. Nieto-Vesperinas, "Experimental two-dimensional field mapping of total internal reflection lateral beam shift in a self-collimated photonic crystal," Applied Physics Letters, vol. 94, p. 061121, 2009.
[13] S.-Y. Su and T. Yoshie, "Optical surface edge Bloch modes: low-loss subwavelength-scale two-dimensional light localization," Optics letters, vol. 37, pp. 4398-4400, 2012.
[14] S. Dyakov, A. Baldycheva, T. Perova, G. Li, E. Astrova, N. Gippius, S. Tikhodeev, "Surface states in the optical spectra of two-dimensional photonic crystals with various surface terminations," Physical Review B, vol. 86, p. 115126, 2012.
[15] R. Moussa, B. Wang, G. Tuttle, T. Koschny, and C. Soukoulis, "Effect of beaming and enhanced transmission in photonic crystals," Physical Review B, vol. 76, p. 235417, 2007.
[16] L.Chen, Y. M. Zhu, X. F. Zang, B. Cai, Z. Li, L. Xie and S. L. Zhuang, "Mode splitting transmission effect of surface wave excitation through a metal hole array",
Light Sci. Appl. 2, e60 (2013).
[17] Y. Cao, J. Hou, and Y. Liu, "Convergence problem of plane-wave expansion method for photonic crystals," Physics Letters A, vol. 327, pp. 247-253, 2004
[18] https://www.lumerical.com/tcad-products/fdtd/
[19] Brenger. Jean-Pierre, "Perfectly matched layer (PML) for computational electromagnetics," Synthesis Lectures on Computational Electromagnetics, vol.2, no.1 pp.1–117, 2007.
[20] Miri. Mehdi, Naqavi. Ali, Khavasi. Amin, Mehrany. Khashayar, Khorasani. Sina and Rashidian. Bizhan, " Geometrical approach in physical understanding of the Goos-H?nchen shift in one-and two-dimensional periodic structures," Optics letters, vol.33, no.24 pp.2940–2942, 2008.
[21] K. Sakoda, "Optical Properties of Photonic Crystal," Springer, Berlin Heidelberg New York, 2001.
[22] S. Foteinopoulou and C. M. Soukoulis, "Electromagnetic wave propagation in two-

dimensional photonic crystals: a study of anomalous refractive effects," Physical Review B, vol.72, no.16 pp.165112, 2005.

[23] M. Notomi, "Theory of light propagation in strongly modulated photonic crystals: Refractionlike behavior in the vicinity of the photonic band gap," Physical Review B, vol.62, no.16 pp.10696–10705, 2000.

[24] J. Joannopoulos, S. Johnson, J. Winn and R. Meade, "Photonic crystal: Molding the flow of light," 2nd ed, Princeton University Press, 2008.

[25] S. Foteinopoulou, M. Kafesaki, E. Economou and C. M. Soukoulis, "Backward surface waves at photonic crystals," Physical Review B, vol.75, no.24 pp.245116, 2007.

[26] D. Meade. Robert, D. Brommer. Karl, M. Rappe. Andrew and J. D . Joannopoulos, "Electromagnetic Bloch waves at the surface of a photonic crystal," Physical Review B, vol.44, no.19 pp.10961–10964, 1991.

[27] R. Zia, M. Selker and M. Brongersma, "Leaky and bound modes of surface plasmon waveguides," Physical Review B, vol.71, no.16 pp.165431-165439, 2005.

[28] Z. Ruan, M. Qiu, S. Xiao, S. He and T. Lars, "Coupling between plane waves and Bloch waves in photonic crystals with negative refraction," Physical Review B, vol.71, no.4 pp.045111, 2005.

[29] T˜ orm¨a. P and Barnes. WL, "Strong coupling between surface plasmon polaritons and emitters," arXiv preprint arXiv:1405.1661, 2014.

[30] Gaspar-Armenta. JA and Villa-Villa. F, "Coupling of a 2D photonic crystalCmetal surface wave to photonic crystal waveguide modes," Journal of Optics, vol.16, no.3 pp.035501, 2014.